\documentclass[prl,twocolumn,superscriptaddress,preprintnumbers,amsmath,amssymb,floatfix]{revtex4}
\usepackage{graphicx}

\begin{document}
\title{Majority-spin non-quasiparticle states in half-metallic ferrimagnet Mn$_2$VAl}
\author{L. Chioncel}
\affiliation{Institute of Theoretical Physics, Graz University of Technology,
A-8010 Graz, Austria}
\affiliation{Faculty of Science, University of Oradea, RO-410087 Oradea, Romania}
\author{E. Arrigoni}
\affiliation{Institute of Theoretical Physics, Graz University of Technology,
A-8010 Graz, Austria}
\author{M.I.~Katsnelson}
\affiliation{Institute for Molecules and Materials, Radboud
University of Nijmegen, NL-6525 ED Nijmegen, The Netherlands}
\author{A.I.~Lichtenstein}
\affiliation{Institute of Theoretical Physics, University of Hamburg,  DE-20355 Hamburg, Germany}

\begin{abstract}
The density of non-quasiparticle states in the ferrimagnetic full-Heuslers Mn$_2$VAl alloy
is calculated from first principles upon appropriate inclusion of correlations. In contrast
to most half-metallic compounds, this material displays an energy gap in the majority-spin
spectrum. For this situation, non-quasiparticle states are located below the Fermi level,
and should be detectable by spin-polarized photoemission. This opens a new way to study
many-body effects in spintronic-related materials.
\end{abstract}

\maketitle

Half metals display a particular type of itinerant-electron
magnetism as well as unusual electronic properties: they are metallic
for one spin channel, and insulating or semiconducting for the
opposite one \cite{gr.mu.83,ka.ir.08}. Electronic structure calculations
based on density functional theory offer an explanation for
the half-metallicity based on the interplay between the crystal structure,
the valence electron count, the covalent bonding, and the large exchange
splitting in addition to symmetry constrains.
The expected $100\%$ spin polarization of half-metals turned out to be an excellent
motivation in developing the field of spintronics both from a
theoretical and an experimental point of view \cite{zu.fa.04,ka.ir.08}.
In reality many potential half-metallic ferromagnets exhibit a dramatic decrease of
bulk spin polarization at temperatures well below their Curie temperature.
In order to understand such a behavior from a theoretical point of view it is
necessary to consider finite temperature many-body effects
\cite{ka.ir.08}.

An important effect of dynamical electron correlations in
half-metals is the existence of non-quasiparticle (NQP) states
\cite{ed.he.73,ir.ka.90,ir.ka.94}. These states  contribute significantly
in reducing the tunneling transport in heterostructures containing HMF
\cite{ir.ka.02,mc.fa.03,tk.mc.01,mc.fa.02,ir.ka.06}, even in the presence
of disorder. NQP states strongly influence the value and temperature
dependence of the spin polarization in HMF
\cite{ch.sa.08,ir.ka.94,ka.ir.08}, which is of primary
interest for potential applications. These states originate from spin-polaron
processes whereby the minority spin low-energy electron excitations, which
are forbidden for HMF in the single-particle picture, are possible
as superpositions of majority-spin electron excitations and virtual
magnons \cite{ed.he.73,ir.ka.90,ir.ka.94,ka.ir.08}. Recently
we have applied the LDA+DMFT (local density approximation plus
dynamical mean field theory) method (for review of this approach,
see Ref.\onlinecite{ko.sa.06}) to describe from first principles
the non-quasiparticle states in several half-metals
\cite{ch.ka.03,ch.ka.05,ch.ar.06,ch.ma.06,ch.al.07}.
Up to now, our studies were restricted to half metals with a gap in
the {\it minority} spin channel. In this situation NQP states
appear just {\it above} the Fermi level \cite{ir.ka.94}.

On the contrary, it was predicted that in half-metallic materials
with a gap in the majority (say, ``up'') spin channel, NQP states should
appear below the Fermi level~\cite{ed.he.73,ir.ka.90,ir.ka.94}.
This asymmetry can be understood
in terms of electron-magnon scattering processes,
as presented in the followings.

A well studied model which takes into account the interaction of
charge carriers with local moments is the {\it s-d} exchange
model. The interacting part of the Hamiltonian is given by
$-I \sum_{}{\bf S}_i {\bf \sigma}_{\alpha \beta}c_{i\alpha}^{\dagger} c_{i \beta}$, 
where $I$ is the {\it s-d}
exchange parameter, ${\bf S}_i$ represents the localized spin
operators, ${\bf \sigma}_{\alpha \beta}$ are the Pauli matrices,
and $c_{i \sigma}$ are operators for conduction electrons. The NQP
picture turns out to be essentially different for the two possible
signs of the $s-d$ exchange parameter.

The ground state of the system with $I>0$ (assuming that the Fermi
energy is smaller than the spin splitting $2IS$) has maximum spin
projection and, thus, the minority-electron band should be empty.
For this case ($I>0$), NQP states in
the minority spin gap develop as a superposition of the
majority-electron states plus magnon states, and of the
minority-electron states, so that the {\it total} spin projection
of the system is conserved. As a result of this spin-polaronic
effect, the minority-electron density of states has a tail
corresponding to the virtual conduction electron spin-flip
processes with magnon emission. However, these virtual flips are
impossible below the Fermi energy $E_F$ due to the Pauli principle
(all majority-electron states are already occupied and thus
unavailable). Therefore, for the positive s-d exchange interaction
the NQP states form above $E_F$.

Contrary, for negative $I$ the minority-spin band lies below the
majority-spin one \cite{ka.ir.08,ir.ka.94}. Occupied minority-spin states 
can be superposed with majority-electron states plus magnons, with 
conserved total spin projection, so the NQP states occur below $E_F$. 
At the same time, for $I<0$ the ferromagnetic ground state is
non-saturated and thus zero-point magnon fluctuations are allowed.
It is the fluctuations which are responsible for formation of {\it
occupied} majority-electron NQP states there.

Formally, the difference between $I<0$ and the previous $I>0$ cases, can be 
explained in terms of a particle-hole transformation $c_{i \sigma}^{\dagger}
\rightarrow d_{i \bar{\sigma}}$, and $c_{i \sigma} \rightarrow
d_{i \bar{\sigma}}^{\dagger}$. This modifies the {\it s-d}
exchange Hamiltonian into
$I \sum_{}{\bf S}_i {\bf \sigma}_{\alpha \beta}d_{i \alpha}^{\dagger}d_{i \beta}$. 
In other words, the Hamiltonian with $I>0$
for electrons is equivalent to that with $I<0$ for holes.

The above argument based on the {\it s-d} exchange model can
be generalized for arbitrary multi-band half-metallic electronic
structures \cite{ir.ka.94,ka.ir.08}. The conclusion remain unchanged:
{\it for the case of minority-electron gap, NQP states
are situated above the Fermi energy, while for the cases when the gap is
present for majority-electrons, NQP states are formed below the Fermi energy.}

Most HMF materials have a gap in the minority spin channel
so that NQP states arise above the Fermi level.
As a consequence, these states cannot be studied
by the very well-developed and accurate technique
of spin-polarized photoemission \cite{john.97},
which can only probe occupied states.
The spin-polarized Bremsstrahlung Isochromat Spectroscopy (BIS)
probing unoccupied states \cite{dona.94} has a much lower
resolution. For this reason, HMF with a gap in the majority-spin channel,
and, consequently, NQP states in the occupied region of the spectrum,
allow for a detailed experimental analysis of these correlation-induced
states and are, therefore, potentially of great interest. It is the purpose
of the present work to perform an electronic structure calculation
based on a combination of the generalized-gradient approximation (GGA)
and  of DMFT for the half-metallic ferrimagnetic full-Heusler alloy Mn$_2$VAl,
which has a gap in the majority spin channel. By appropriately taking into
account effects due to electronic correlations, we demonstrate explicitly the
existence of majority spin NQP states arising just below the Fermi level, and
study the temperature dependence of their spectral weight.

In full Heusler compounds with the formula X$_2$YZ,  Mn atoms
usually occupy the Y-position, while  compounds in which Mn assumes
the X-position Mn$_2$YZ, are very rare. The prototype from the latter
category is Mn$_2$VAl,  for which a large number of theoretical
and experimental investigations have been made. Neutron diffraction
experiments \cite{it.na.83} demonstrated the existence of a ferrimagnetic state in
which Mn has a magnetic moment of $1.5 \pm 0.3 \mu_B$ and V moment
is $-0.9 \mu_B$. X-ray diffraction and magnetization measurements
\cite{ji.ve.01} found a total magnetic moment of $1.94 \mu_B$ at
$5K$, close to the half-metallic value of $2 \mu_B$. The Curie
temperature of the sample was found to be about $760K$ and the
loss of half-metallic character was attributed to the small amount
of disorder. Electronic-structure calculations performed by
Ishida \cite{is.as.84} within the local-density approximation (LDA),
predict the ground state of Mn$_2$VAl to be close to half-metallicity.
Weht and Pickett \cite{we.pi.99} used the GGA for the exchange correlation
potential and showed that Mn$_2$VAl is a half-metallic ferrimagnet with
atomic moments in very good agreement with the experiment. Recent
calculations of the exchange parameters for Mn$_2$VAl~\cite{sa.sa.05}
show a strong $Mn-V$ exchange interaction that influence the ordering
in the Mn sublattice. The estimated Curie temperatures are in good
agreement with the experimental values~\cite{sa.sa.05}.
The intermixing between V and Al atoms in the Mn$_2$VAl alloy
showed that a small degree of
disorder decreases the spin polarization at the Fermi level
from its ideal $100\%$ value, but the resulting alloy
Mn$_2$V$_{1-x}$Al$_{1+x}$ still show an almost half-metallic
behavior \cite{oz.ga.06,ga.oz.07}.

According to the ideal full Heusler ($L2_1$) structure, the
V atom occupy the $(0,0,0)$ position, the Mn atoms are situated
at $(1/4,1/4,1/4)a$ and $(3/4,3/4,3/4)a$, and the Al
at $(1/2,1/2,1/2)a$, where $a=5.875\AA$ is the lattice constant
of the Mn$_2$VAl compound.
In our work,
correlation effects in the valence V and Mn $d$ orbitals are included
via an on-site electron-electron interaction in the form
$\frac{1}{2}\sum_{{i \{m, \sigma \} }} U_{mm'm''m'''}
c^{\dag}_{im\sigma}c^{\dag}_{im'\sigma'}c_{im'''\sigma'}c_{im''\sigma} $.
The interaction is treated in the framework of dynamical mean field theory (DMFT)
\cite{ko.sa.06}, with a spin-polarized T-matrix Fluctuation Exchange (SPTF) type
of impurity solver \cite{ka.li.02}. 
Here, $c_{im\sigma}/c^\dagger_{im\sigma}$
destroys/creates an electron with spin $\sigma$ on orbital $m$ on
lattice site $i$.
The Coulomb matrix elements $U_{mm'm''m'''}$ are expressed in the usual
way~\cite{im.fu.98} in terms of  three Kanamori parameters $U$, $U'=U-2J$
and $J$. Typical values for Coulomb ($U=2$eV) and Stoner ($J=0.93$eV) parameters
were used for Mn and V atoms.  The above value of U is considerably smaller 
than the bandwidth of Mn$_2$VAl (7--8eV) therefore 
the use of a 
perturbative SPTF-solver  
is justified. In addition, the same solver was used to investigate 
spectroscopic properties of transition metals with remarkable results 
\cite{mi.ch.05,mi.eb.05,br.mi.06,ch.mi.06,gr.ma.07}.  

Since the static contribution from correlations is already
included in the local spin-density approximation (LSDA/GGA), so-called
``double counted'' terms must be subtracted. To achieve this, we replace
$\Sigma_{\sigma}(E)$ with $\Sigma_{\sigma}(E)-\Sigma_{\sigma}(0)$
\cite{li.ka.01} in all equations of the DMFT procedure \cite{ko.sa.06}.
Physically, this is related to the fact that DMFT only adds {\it dynamical}
correlations to the LSDA/GGA result. For this reason, it is believed that this kind
of double-counting subtraction ``$\Sigma(0)$'' is more appropriate for a DMFT
treatment of metals than the alternative static ``Hartree-Fock'' (HF) subtraction
\cite{pe.ma.03}.

\begin{figure}[h]
\includegraphics[width=1.00\columnwidth]{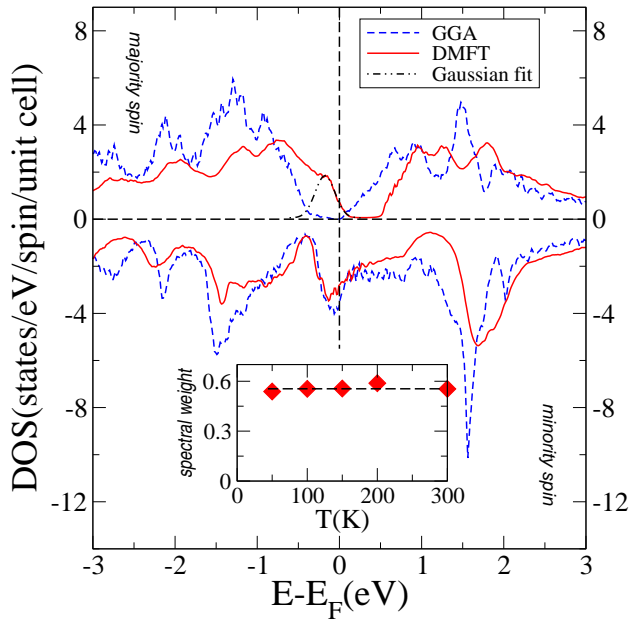}
\caption{(color online) Total density of states,
computed within  GGA (dashed/blue), and GGA+DMFT (full/red).
The Gaussian fit to the density of NQP states is shown as a dotted-dashed 
(black) line just below the Fermi level for the majority spin channel. The 
temperature dependent spectral weight of NQP states is displayed in the inset.}
\label{dos_nqp_mn2val}
\end{figure}

In Fig. \ref{dos_nqp_mn2val} we present the total density of states computed
in GGA and GGA+DMFT, for T=200K. The GGA density of states  displays a gap
of about 0.4eV in the majority spin channel  in agreement with previous 
calculations~\cite{we.pi.99}. As expected from the s-d model calculation, 
majority spin NQP states are visible just below the Fermi level, with a peak 
around $-0.25eV$. In order to evaluate the spectral weight of these NQP states, 
we fit the low-energy density of states below the Fermi level in the majority 
channel with a Gaussian centered around the peak position. The NQP spectral 
weight is then defined as the area below the Gaussian curve.
The inset shows the NQP spectral weight for several temperatures up to 300K.
It is interesting to note that within the computed temperature range $50 \le T \le 300$
these values are almost constant and considerably larger in comparison with
similar values for (NiFe)MnSb \cite{ch.ar.06}.
The data presented in the inset can be extrapolated down to $T=0K$, and a
spectral weight of $\approx 0.544 \pm 0.018(states/Mn-d)$ is obtained. This demonstrates that
NQP states are present also at T=0K, and obviously they
are not captured by the mean-field, GGA result. As it will be discussed 
below,  NQP states predominantly consist of {\it Mn-d} electrons, so the 
value for the integrated spectral weight (inset of Fig. \ref{dos_nqp_mn2val}) 
only comes from {\it Mn-d} orbtials.

\begin{figure}[h]
\includegraphics[width=1.00\columnwidth]{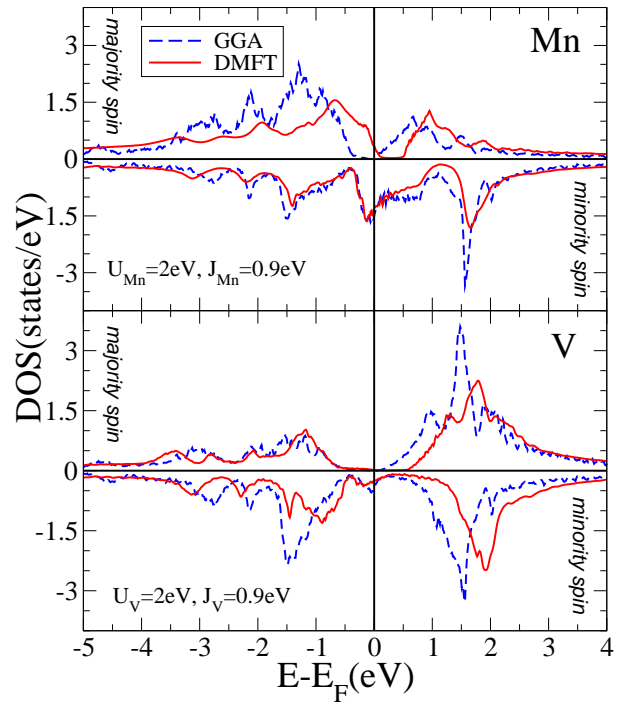}
\caption{(color online)
Atom resolved density of states, computed within the GGA (dashed/blue) and GGA+DMFT
(full/red) approach, at T=200K. The majority spin NQP states
are visible in the Mn-3d$^\uparrow$ DOS just below the Fermi level.}
\label{dos_mn2val}
\end{figure}

The atom resolved DOS is presented in Fig. \ref{dos_mn2val}. In GGA, the net magnetic
moment per unit cell is $2\mu_B$ with parallel Mn moments having values close
to $1.6 \mu_B$ and oppositely oriented V moments close to $-0.8\mu_B$.
Below the gap, the majority spin total DOS is mainly of Mn character. The Mn and
V moments have a  strong $t_{2g}$ character, and a small Al contribution 
to the magnetic moment is present. Most
of the $V$ majority spin states lie above the gap, along with the Mn $e_{g}$ states.
Minority-spin states below 0.5eV have roughly an equal amounts of $V(t_{2g})$ and
Mn character. States around the Fermi energy have a predominant $Mn(t_{2g})$ character,
in agreement with~\cite{we.pi.99}.
In contrast to the GGA results, the many-body DMFT calculation
(see Fig. \ref{dos_mn2val}) yields a significant DOS for the
majority spin states just below the Fermi level. These are the
{\it majority spin NQP states} discussed above
\cite{ch.ka.03,ch.ar.06,ch.sa.08,ch.ka.05,ch.ma.06,ch.al.07}.
As can be seen in Fig. \ref{dos_mn2val} majority spin NQP states
are predominantly of $Mn-3d^{\uparrow}$ character. Their spectral
weight is quite significant (see inset of Fig. \ref{dos_nqp_mn2val})
so that accurate spin-polarized photoemission experiments should be able
to identify the existence of such states. In contrast, majority spin
V(t$_{2g}$)  states below the Fermi energy are not significantly changed.
Above the Fermi level, the Mn(e$_g$) and V(e$_g$) states are pushed to
higher energy, such that a gap is formed just above $E_F$. In the minority
spin channel below $E_F$, both V and Mn(t$_{2g}$) states are slightly modified,
while above $E_F$, V(e$_g$) states are shifted to higher energies by $0.5eV$.
Around the Fermi level, the dominant $Mn-3d^{\downarrow}$ DOS is not
significantly changed with respect to the GGA values.

The applicability of the local DMFT approach to the problem of the existence of 
NQP states has been discussed in ref. \cite{ka.ir.08} and \cite{ch.ka.03}.
It is essential to
stress that the accurate description of the magnon spectrum is not important
for the existence of nonquasiparticle states and for the proper estimation
of their spectral weight, but can be important to describe an explicit shape of
the density of states ``tail'' in a very close vicinity of the Fermi energy.

\begin{figure}[h]
\includegraphics[width=1.0\columnwidth]{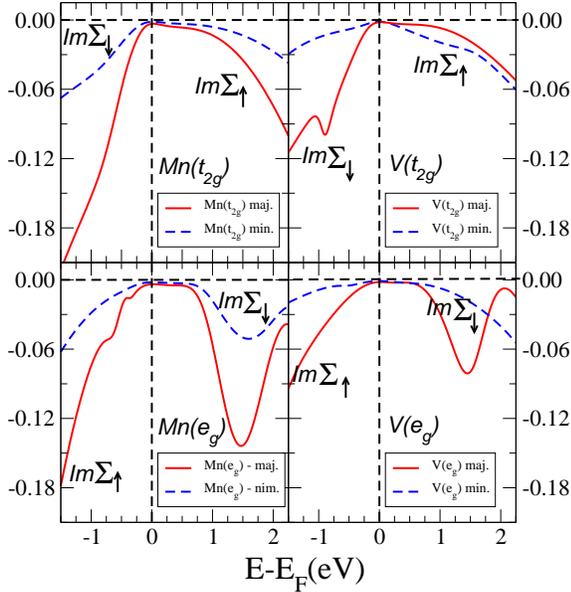}
\caption{(color online) 
Imaginary part of the self-energies 
$Im\Sigma_{Mn/V}^{\sigma}$ for
$t_{2g}$-orbitals on Mn (left upper panel) and V (right upper panel).
The solid (red) line shows results for the 
majority($\uparrow$) spins, while the dashed (blue) line for minority spins.
The lower pannels show the corresponding  $Im\Sigma_{Mn/V}^{\sigma}$ for the
$e_g$ orbitals.}
\label{SIGMA_MN_V}
\end{figure}

The imaginary part of the atom and orbital resolved self-energies, for T=200K 
are presented in Fig. \ref{SIGMA_MN_V}. For the minority Mn-$e_g$, V-$e_g$ and 
V-$t_{2g}$-orbitals we observe that the imaginary part of the self-energy has a rather 
symmetric energy dependence around the Fermi level, with a normal
Fermi-liquid-type behavior 
$-Im\Sigma_{Mn/V}^{\downarrow}(E) \propto (E-E_F)^2$. The majority spin 
$-Im\Sigma_{Mn/V}^{\uparrow}(E)$
shows a significant increase right below  the Fermi level which is more pronounced for
the $t_{2g}$-orbitals. In addition, a slight kink is evidenced  for an energy around 
-0.25eV. The majority-state nonquasiparticles are visible in the majority spin channel 
Fig. \ref{dos_mn2val} at about the same energy. 
These results shown in
Fig. \ref{SIGMA_MN_V} suggests
that many-body effects are stronger on Mn than on V sites.  
Therefore, NQP states are mainly 
determined by the  {\it Mn-d} atoms.

The behavior of the imaginary part 
of the self-energy (Fig. \ref{SIGMA_MN_V}) and the Green function (Fig. \ref{dos_mn2val})
can be correlated with the analysis of the spin-resolved optical conductivity.
We have estimated the latter for different temperatures
whithin the GGA and GGA+DMFT approaches using an
approximation of constant matrix elements. 
Already at 50K the majority-spin optical spectra shows the appearance of a Drude peak
signaling the closure of the majority
spin gap.
With increasing 
temperatures, spectral weight is transfered towards the Drude peak 
contributing to the 
depolarization discussed below in Fig. \ref{pol_mag}. 

As it was demonstrated previously \cite{ed.he.73,ir.ka.90,ir.ka.94,ka.ir.08},
the nonquasiparticle
spectral weight in the density of states (Fig. \ref{dos_mn2val}) 
is proportional to the   
imaginary part of the self-energy (Fig. \ref{SIGMA_MN_V}), therefore it is
determined by the  quasiparticle decay, which is the reason for the
name of these states.

\begin{figure}[h]
\includegraphics[width=0.99\columnwidth]{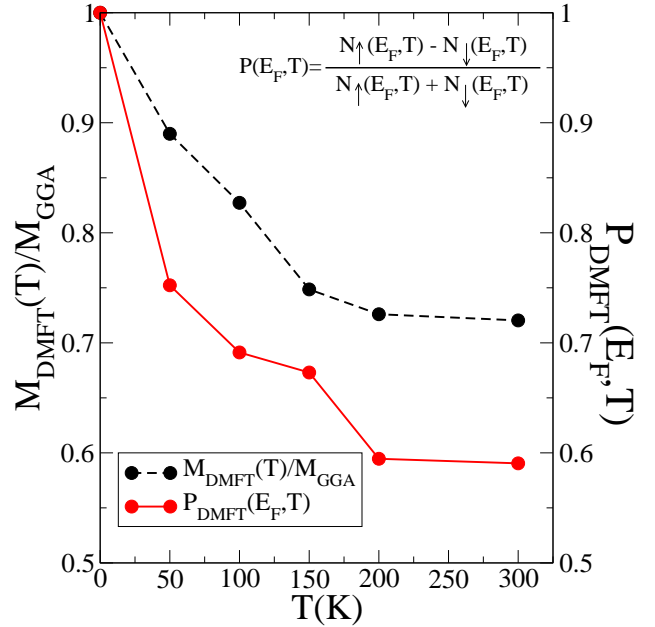}
\caption{(color online) Temperature dependence of the spin polarization of
conduction electrons (full/red) at the Fermi level $P(E_F,T)$, and normalized
magnetization $M(T)/M(0)$ (dashed/black).}
\label{pol_mag}
\end{figure}

Fig. \ref{pol_mag} displays the temperature dependence of the magnetization
obtained directly from the GGA+DMFT calculations and the spin polarization
at the Fermi level, obtained using the relation
$P(E_F,T)=(N_\uparrow(E_F,T)-N_\downarrow(E_F,T))/(N_\uparrow(E_F,T)+N_\downarrow(E_F,T))$,
$N_{\sigma}$ being the density of states. These results reflect
a general trend valid for half-metals in the presence of NQP states
\cite{ka.ir.08,ir.ka.06,ch.ka.03,ch.ka.05,ch.ar.06,ch.ma.06,ch.al.07},
namely that
magnetization and polarization  behave differently as function of temperature.
However, as can be seen in Fig.~\ref{pol_mag}, this difference is not
as sharp as in other HMF materials \cite{ch.sa.08,ch.al.07}.

The effect of disorder on half-metallicity was recently discussed in
Mn$_2$V$_{1-x}$Al$_{1+x}$ alloys, for $-0.2 \le x \le 0.2$ \cite{oz.ga.06,ga.oz.07,ji.ve.01}.
The excess of both Al and V atoms ($x=0.1/-0.1$ or $x=0.2/-0.2$) has the effect of
shrinking the gap to zero, but with the
Fermi level situated within the gap.
In addition, the Mn moment is not affected by disorder and remains
constant, in contrast to the V moment. Spin polarization is decreased
by about $10\%$, with electrons around the Fermi
level having a dominant minority spin character \cite{oz.ga.06}.
In contrast, many-body correlations have a much more dramatic effect.
For non-zero temperatures, all atomic magnetizations are decreased. For
instance, near room temperature ($T=300K$) the strongest decrease occurs
in V, for which the moment drops almost by $47\%$, the Mn moment is reduced
by $32\%$, while the Al experiences just a small reduction by $4\%$.
As one can see from Fig. \ref{pol_mag},
already at $50K$ polarization drops to $75\%$, and is further decreased
upon increasing the temperature. As we discussed previously for the case of
FeMnSb \cite{ch.ar.06}, many-body induced depolarization is significantly
stronger than the effect of disorder or of other spin-mixing mechanisms
such as spin-orbit coupling. This observation seems to hold also for the
case of Mn$_2$VAl, although we can not exclude the fact that for a larger
degree of disorder, the material could possibly depart from its almost
half-metallic situation displayed for small degree of substitution ($-0.2<x<0.2$).

%\section{Conclusion}

In conclusion, in this paper we have shown for a specific material
that NQP states are also present in half-metals with a gap in the
majority spin channel, and appear just below the Fermi level, as
predicted in model calculations~\cite{ir.ka.90}.
In the case of Mn$_2$VAl, these states mainly consist of
{\it Mn-3d}$^\uparrow$ electrons and have a considerable spectral weight.
Although this material was reported to be a half-metal from electronic structure
calculations \cite{we.pi.99}, the experimental evidence is not
clear. Several reasons are invoked such as existence of defects or the reduced
symmetry at surface and interfaces. From a theoretical point of view, we show that
correlation-induced NQP states significantly change the majority spin
electronic states, thus reducing the spin polarization at $E_F$.
The appearance of NQP states and its connection with tunneling
magnetoresistance was  recently studied in Co$_2$MnSi-based tunnel magnetic
junction~\cite{ch.sa.08}. A great challenge would be to produce TMR
junctions based on the ferri-magnetic Mn$_2$VAl. This would allow for
a  direct experimental investigation of the existence of majority spin
NQP states. Promising candidate HMF materials with a majority spin gap
of similar magnitude as Mn$_2$VAl are the double perovskites Sr$_2$FeMO$_6$
(M=Mo,Re) or Sr$_2$CrReO$_6$ often associated with collosal magnetoresistance behavior.
In particular the electronic structure of Sr$_2$CrRe$O_6$ shows a closure of
its majority spin gap in the presence of spin-orbit coupling \cite{va.ka.05},
with states having a small spectral weight symmetrically distributed
around the Fermi energy.
Discrepancies between the experiment and theoretical computations were
explained based on possible anti-site disorder \cite{va.ka.05}. We suggest
that a significant density of NQP states could be present in the above
perovskites as well. Work on these lines is in progress.

L.C. and E.A. acknowledge financial support by the Austrian science
fund under project nr. FWF P18505-N16. L.C. also acknowledge the financial 
support offered by Romanian Grant CNCSIS/ID672/2009. M.I.K. acknowledges 
financial support from FOM (The Netherlands).
A.I.L. acknowledge financial support from the DFG (Grants No. SFB 668-A3).

\bibliography{references_database}
\end{document}